\documentclass[fleqn,10pt]{wlscirep}
\usepackage[utf8]{inputenc}
\usepackage[T1]{fontenc}
\usepackage{booktabs}

\title{Pulsed thermal annealing enables switching of chiral antiferromagnetic order with a sub-millitesla field in Mn$_3$Sn}

\author[1,*]{Xiaokang Li}
\author[1]{Jing Zhang}
\author[1]{Xiaodong Guo}
\author[1,*]{Zengwei Zhu}
\affil[1]{Wuhan National High Magnetic Field Center and School of Physics, Huazhong University of Science and Technology, Wuhan, 430074, China}
\affil[*]{lixiaokang@hust.edu.cn, zengwei.zhu@hust.edu.cn, Xiaokang Li and Jing Zhang contributed equally to this work.}

\begin{abstract}
The manipulation of antiferromagnetic (AFM) order is a central theme in modern spintronics. In this work, we achieve reliable switching of the chiral AFM state in the Weyl antiferromagnet Mn$_3$Sn using a heat pulse combined with a very small magnetic field as small as 0.1~mT. By systematically measuring the anomalous Hall effect (AHE) in high-quality single crystals, we show that the field needed for switching decreases as the temperature approaches the N\'eel temperature $T_N$, and vanishes at $T_N$. Pulsed thermal annealing above $T_N$ followed by cooling in a tiny external field enables full and reproducible switching of the magnetic octupole order. Our results show that thermal softening (heating above $T_N$ to temporarily remove the magnetic anisotropy) is a key step that lowers the energy barrier to nearly zero. This allows an extremely weak directional field (like the effective field from spin–orbit torque in thin-film devices) to set the final magnetic state during cooling. We also provide a simple model to estimate the temperature rise in nanoscale devices under current pulses, giving practical guidance for device design. This work highlights that thermal effects are not a side issue but an important partner to spin torques, and suggests that future work should take both into account.
\end{abstract}

\begin{document}

\flushbottom
\maketitle
\thispagestyle{empty}

\section*{Introduction}

The rapid advancement of nanofabrication has pushed conventional semiconductor memory devices toward their physical limits, spurring the search for alternative information storage paradigms. Spintronics, which exploits both the charge and spin of electrons, has emerged as a promising candidate due to its non-volatility, high speed, and compatibility with existing electronic infrastructures~\cite{Baibich1988, Slonczewski1996}. However, traditional ferromagnetic spintronic devices still face challenges such as stray fields and relatively slow dynamics. Antiferromagnetic (AFM) materials offer an attractive alternative because of their ultrafast spin dynamics (terahertz scale), absence of stray fields, and inherent robustness against external magnetic perturbations~\cite{Jungwirth2016, Baltz2018, Jungwirth2018, Vsmejkal2018}.

Among them, the chiral AFM Mn$_3$Sn has emerged as a particularly compelling candidate~\cite{Chen2014, Kubler2014, Nakatsuji2015}. Its non-collinear $120^\circ$ spin structure on a kagome lattice gives rise to a cluster magnetic octupole order that breaks time-reversal symmetry~\cite{Suzuki2017}. This symmetry breaking generates a large anomalous Hall effect (AHE) at room temperature without net magnetization~\cite{Nakatsuji2015}, as well as a series of other anomalous Hall transport phenomena, including the anomalous Nernst effect~\cite{Ikhlas2017, Li2017}, thermal Hall effect~\cite{Li2017, Xu2020}, magneto-optical Kerr effect~\cite{Higo2018, Balk2019}, and topological and planar Hall effects~\cite{Li2018, Rout2019, Li2019, Xu2020b}. Importantly, the octupole vector $\mathbf{K}$ can be electrically switched, and its orientation is read out via the AHE, making Mn$_3$Sn a promising building block for all-antiferromagnetic memory and logic devices~\cite{Tsai2020, Takeuchi2021, Higo2022, Yoon2023, Chen2023, Takeuchi2025, Ogawa2026, Zhou2026}.

A crucial yet often overlooked aspect of switching in Mn$_3$Sn is thermal softening~\cite{Krishnaswamy2022}, as the applied current density is so large (10$^7$ A/cm$^2$) that its heating effect cannot be neglected. As the temperature approaches the Néel temperature $T_N \approx 425$~K, the magnetic anisotropy that pins the octupole order vanishes. Heating above $T_N$ erases the magnetic order and eliminates the reorientation barrier. Upon cooling, even a weak directional stimulus—a small magnetic field or the effective field from a spin current—can dictate the final orientation of $\mathbf{K}$. Despite its physical clarity and practical importance, thermal softening has not been fully discussed in many recent studies, particularly those on spin–torque-driven switching~\cite{Tsai2020, Takeuchi2021, Higo2022, Yoon2023, Chen2023, Takeuchi2025, Ogawa2026, Zhou2026}. Our work aims to fill this gap.

In this work, we directly isolate and demonstrate the role of thermal softening through controlled pulsed thermal annealing experiments on bulk single-crystal Mn$_3$Sn. Using an external heater and a thermocouple, we independently control temperature and magnetic field, thereby completely decoupling thermal effects from spin-torque effects. We show that heating the sample above $T_N$ enables a magnetic field as small as 0.1 mT to achieve near-complete switching of the octupole order upon cooling.  Based on these findings, we argue that thermal softening is not an undesirable side effect, but rather a crucial step that lowers the energy barrier, allowing even very weak directional fields—such as those generated by spin–orbit torque—to accomplish deterministic switching. Finally, we provide a simple analytical model to estimate the transient temperature rise in thin-film devices under current pulses, offering practical guidance for incorporating thermal softening into future device design.

\section*{Experimental results}

\begin{figure}[ht]
\centering
\includegraphics[width=0.8\linewidth]{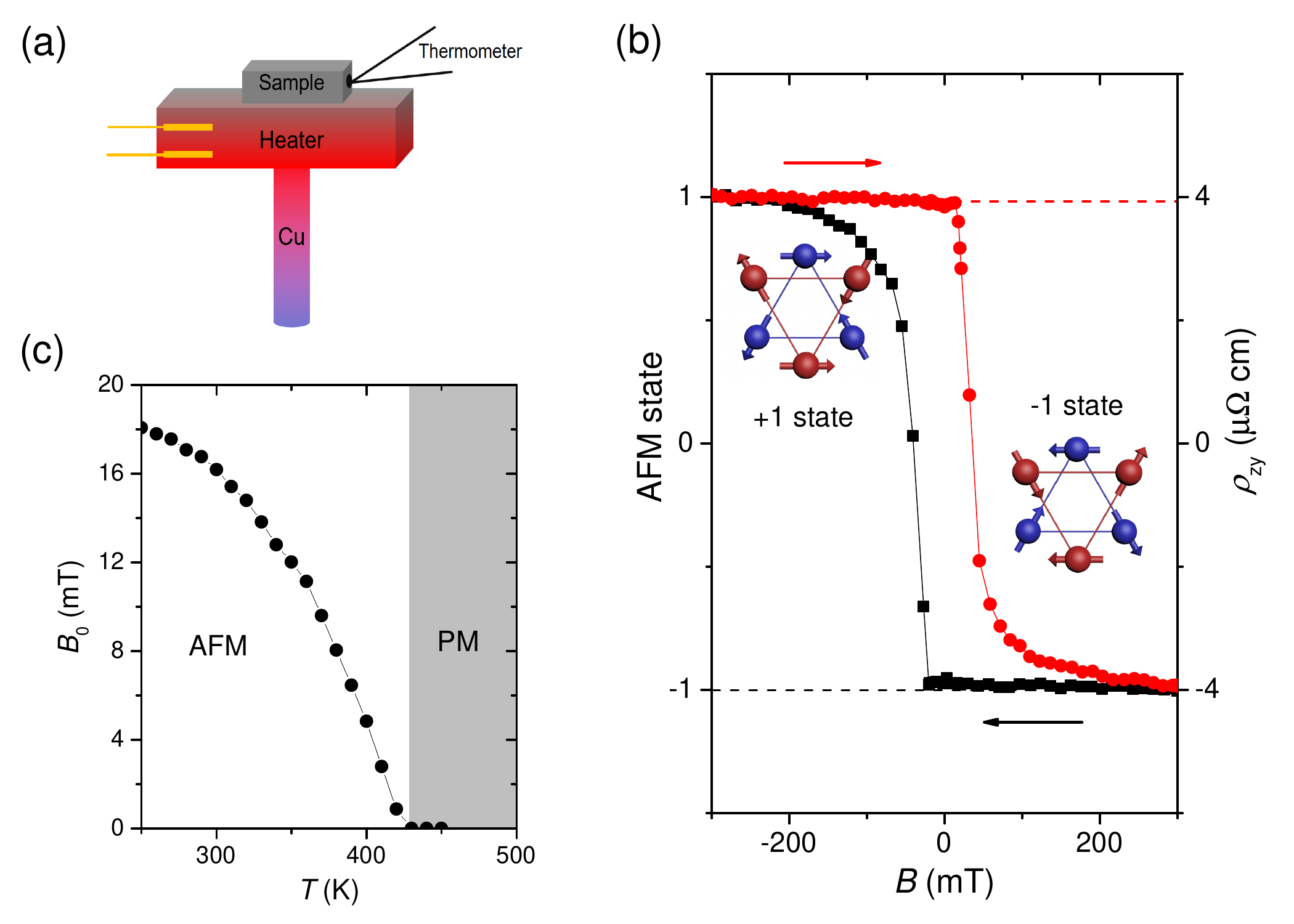}
\caption{\textbf{Setup, chiral antiferromagnetic order and threshold field.} (a) Schematic of the pulsed thermal annealing setup. (b) Field dependence of Hall resistivity at 250~K showing the AHE hysteresis, with the two Hall plateaus representing the positive and negative chiral antiferromagnetic states, respectively. (c) Temperature dependence of the threshold field $B_0$ (the minimum magnetic field required to switch the octupole order at a given temperature), vanishing at $T_N$ around 425 K.}
\label{fig:setup}
\end{figure}

High-quality single crystals of Mn$_3$Sn were grown using the Bridgman–Stockbarger method. 
The Hall resistivity $\rho_{zy}$ was measured using a standard four-probe configuration with the magnetic field applied along the $x$ direction (in-plane) and the current along the $z$ direction (out-of-plane). To achieve independent thermal control, the sample was mounted on a resistive heater chip using insulating glue, ensuring that the heating current did not pass through the sample. An E-type thermocouple attached directly to the sample monitored the temperature, and a copper cylinder served as a heat sink to facilitate rapid cooling after each heating pulse (see Figure~\ref{fig:setup}a). All measurements were performed in a Physical Property Measurement System (PPMS, Quantum Design) with a base temperature of 250 K, where the anomalous Hall effect (AHE) signal is large and stable. The heating current was supplied by a Keithley 6221, and Hall voltages and thermocouple readings were recorded with a Keithley 2182A nanovoltmeter.

Figure~\ref{fig:setup}b shows the field dependence of $\rho_{zy}$ at 250~K. The curve exhibits a clear AHE hysteresis loop, with a threshold field $B_0 \approx 18$~mT required to switch the octupole order between the $+1$ and $-1$ states (defined as the octupole vector pointing along $+x$ or $-x$, respectively). As the temperature increases, the magnetic anisotropy that gives rise to $B_0$ gradually weakens. We measured $B_0$ as a function of temperature (Figure~\ref{fig:setup}c) and found that it decreases monotonically, extrapolating to zero at $T_N \approx 425$~K. Above this temperature, the AHE vanishes entirely due to the paramagnetic phase transition. This behavior confirms that the energy barrier for switching the chiral antiferromagnetic order is progressively reduced by thermal energy and disappears completely at $T_N$.

\begin{figure}[ht]
\centering
\includegraphics[width=0.9\linewidth]{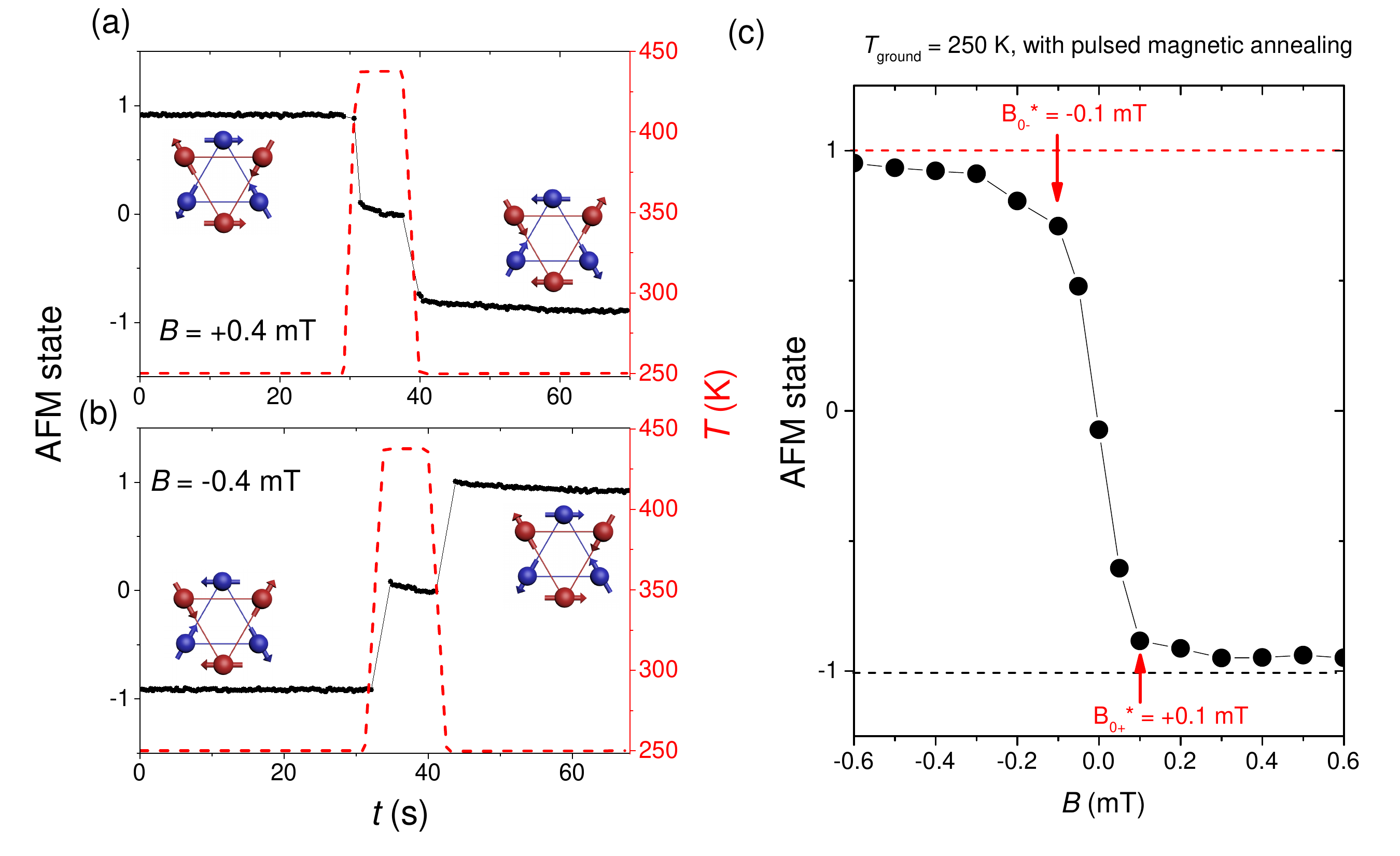}
\caption{\textbf{Pulsed thermal annealing switching.} (a,b) Time evolution of the chiral antiferromagnetic order during heating to 438~K and cooling in (a) +0.4~mT and (b) –0.4~mT. Before each heating pulse, the sample was first magnetized in the opposite direction (e.g., –200~mT for the +0.4~mT cooling run, and +200~mT for the –0.4~mT cooling run), and the field was then slowly ramped to the target cooling field to ensure a well-defined initial state. (c) Switching ratio of the chiral antiferromagnetic order after annealing as a function of the cooling field, exhibiting a sharp transition with no hysteresis. The switching ratio is defined as the AHE signal after annealing normalized by the saturated AHE signal (i.e., the ratio of the measured AHE to the maximum AHE value).}
\label{fig:annealing}
\end{figure}

We then performed pulsed thermal annealing experiments. The sample was heated to 438~K (above $T_N$) for 10~s to ensure thermal equilibrium. This relatively long heating time is necessary because we use an external heater with thermal contact resistance, leading to a long thermal relaxation time. Moreover, the bulk single-crystal nature of our sample requires more time to achieve thermal stability. The sample was then cooled in the presence of a small static magnetic field. Figure~\ref{fig:annealing}a-b shows the time evolution of the chiral antiferromagnetic order (probed by the AHE) during a heating pulse at +0.4~mT and –0.4~mT, respectively. In both cases, the chiral antiferromagnetic order is completely suppressed during the heating plateau (the sample is paramagnetic above $T_N$) and recovers upon cooling, with the sign of the recovered order matching that of the applied field. This demonstrates reliable switching of the octupole order by the tiny cooling field.

We systematically varied the cooling field and, after each annealing cycle, measured the resulting chiral antiferromagnetic order (quantified by the AHE signal). The extracted switching ratio as a function of the cooling field is shown in Figure~\ref{fig:annealing}c. The switching ratio curve is very sharp: a field of only $\pm0.3$~mT yields nearly full saturation ($>90\%$ of the maximum AHE), and even at $\pm0.1$~mT the switching ratio reaches $\sim70\%$. This indicates that the chiral antiferromagnetic order is highly sensitive to the direction of an extremely weak magnetic field during cooling, provided the sample has been heated above $T_N$.

Crucially, the curve shows no hysteresis when cycling the field from 0 to +0.6~mT, down to –0.6~mT, and back to 0, indicating that there is no memory effect—the final state depends solely on the sign of the field applied during cooling, not on the history of previous cycles. At zero cooling field, the switching ratio is nearly zero, confirming that heating above $T_N$ without a directional stimulus leads to a fully demagnetized, random multidomain state.
 
\begin{figure}[ht]
\centering
\includegraphics[width=0.9\linewidth]{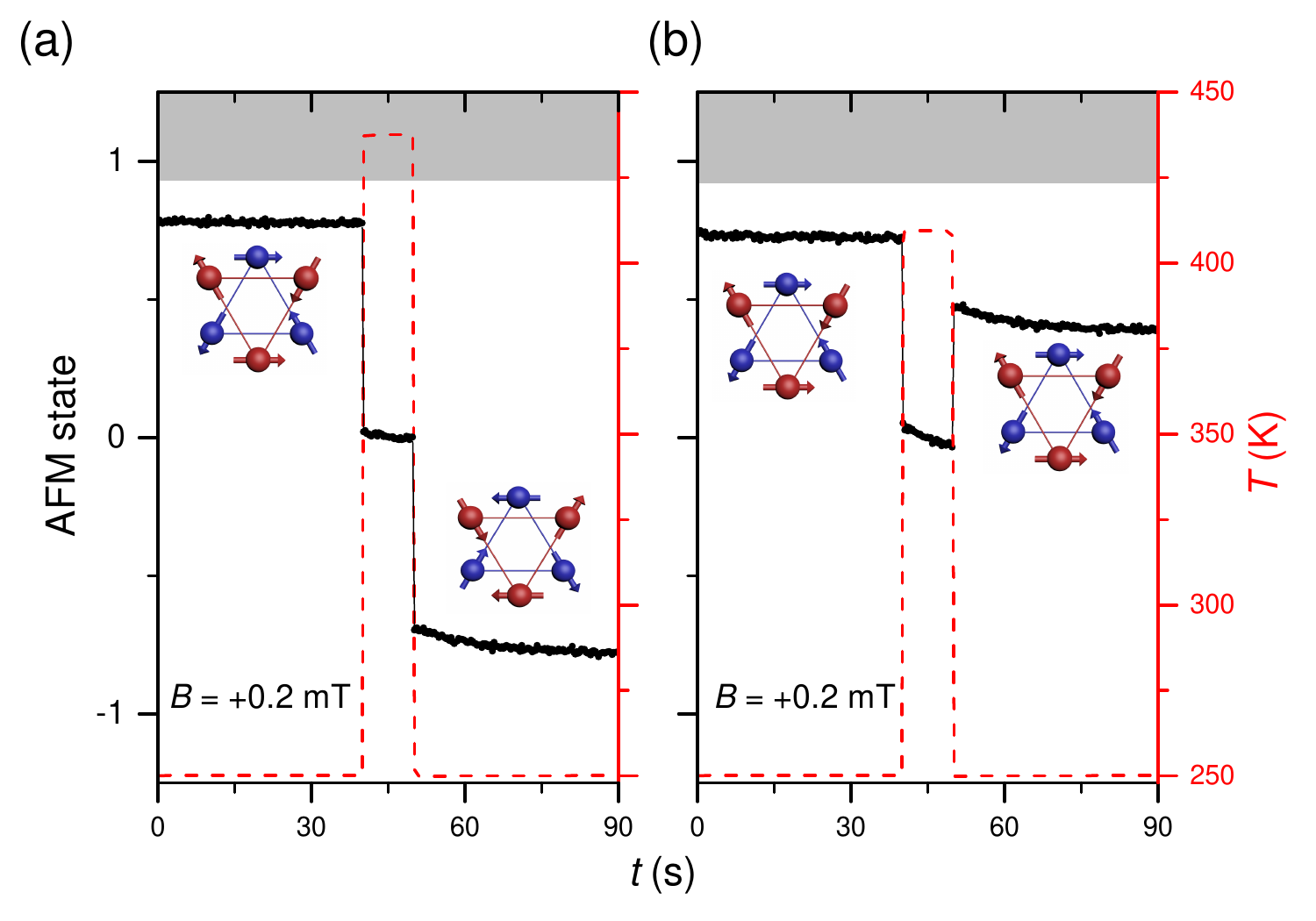}
\caption{\textbf{Crossing $T_N$ is necessary.} (a) Annealing at 438~K (above $T_N$) with a 0.2~mT cooling field: the chiral antiferromagnetic order switches sign. (b) Annealing at 409~K (below $T_N$) with a 0.2~mT cooling field: the chiral antiferromagnetic order does not switch sign. Instead, its original orientation is partially attenuated (as evidenced by a reduced AHE signal) but not reversed.}
\label{fig:crossingT_N}
\end{figure}

To confirm that switching requires heating above $T_N$, we performed comparative experiments at two annealing temperatures: 438~K (above $T_N$) and 409~K (below $T_N$), both with a cooling field of 0.2~mT (Figure~\ref{fig:crossingT_N}a-b). At 438~K, the chiral antiferromagnetic order is completely erased and then switches sign according to the cooling field. At 409~K, the chiral antiferromagnetic order does not switch sign. Instead, its original orientation is partially attenuated (as evidenced by a reduced AHE signal) but not reversed. This result clearly shows that crossing the N\'eel temperature is essential for the weak field to control the final magnetic state. Heating below $T_N$ only partially reduces the existing order (likely due to thermal activation over the reduced barrier) but does not erase it sufficiently to allow a weak field to impose a new orientation.

\section*{Discussion}

Our experiments establish that thermal softening is the essential enabler for deterministic switching of the chiral antiferromagnetic order in Mn$_3$Sn. Heating above $T_N$ eliminates the magnetic anisotropy barrier. Cooling in a tiny directional field (here, as small as 0.1~mT) then selects the final octupole orientation. This two-step process  is fundamentally different from pure spin–orbit torque (SOT) switching, which attempts to rotate the octupole directly against the full anisotropy barrier. 

In thin-film SOT devices, a current pulse simultaneously generates Joule heating and a spin current. If the temperature rise reaches $T_N$, the barrier disappears, and the SOT effective field (typically a few mT) can bias the reorientation during cooling. Thus, SOT should be viewed not as the sole driver, but as the directional bias in a heat-assisted process. This perspective helps to reconcile why some SOT switching experiments required an external magnetic field (when heating was insufficient) while others succeeded with longer pulses or higher current densities (where thermal softening became effective). 

To help design devices that deliberately exploit thermal softening, we provide two simple formulas for the temperature rise $\Delta T$ in the magnetic layer under a current pulse. The formulas assume adiabatic heating (pulse shorter than thermal diffusion out of the active region) and are applicable to two common regimes.

For a continuous thin film, the heating resistance comes mainly from the magnetic layer's bulk resistivity $\rho_e$. The temperature rise is
\[
\Delta T = \frac{J^2 \rho_e \tau}{\rho_m C_p},
\]
where $J$ is current density, $\tau$ pulse duration, $\rho_m$ mass density, and $C_p$ specific heat. For Mn$_3$Sn, $\rho_e \approx 150~\mu\Omega\cdot$cm, $\rho_m \approx 7.5~\text{g/cm}^3$, $C_p \approx 0.4~\text{J/g·K}$. For typical $J \sim 10^7$~A/cm$^2$ and $\tau = 1$~ns, $\Delta T \sim 5$~K. To reach $T_N \approx 425$~K one needs $J \gtrsim 10^8$~A/cm$^2$ or $\tau$ longer than 100~ns.

In nanoscale devices (e.g., magnetic tunnel junctions or nanopillars), the total heating resistance is often dominated by the contact/tunnel resistance $R_c$. Heat is generated in a tiny volume $V_{\text{hot}} \approx A \delta$, where $A$ is the contact area and $\delta$ a characteristic length (e.g., barrier thickness or a few nm). The temperature rise is
\[
\Delta T = \frac{I^2 R_c \tau}{\rho_m C_p V_{\text{hot}}} = \frac{J^2 A R_c \tau}{\rho_m C_p \delta}.
\]
For a typical nanopillar with $R_c = 1$~k$\Omega$, $A = (50~\text{nm})^2$, $J = 10^7$~A/cm$^2$, $\tau = 1$~ns, and $\delta = 5$~nm, we obtain $\Delta T \approx 1000$~K — easily enough to approach $T_N$ with even moderate current densities.

These results provide practical design guidelines: (i) Intentionally engineer the thermal profile so that the device core reaches $T_N$ during the write pulse. (ii) Because the barrier is thereby removed, even a modest SOT effective field can supply the necessary directional bias, reducing power consumption. By embracing thermal softening rather than trying to avoid it, one can achieve low‑power, fast, and reliable all‑electrical switching in chiral antiferromagnets.

\section*{Conclusion}
In this work, we have shown that reliable switching of the chiral antiferromagnetic order in Mn$_3$Sn can be achieved by pulsed thermal annealing: heating the sample above the N\'eel temperature to erase the magnetic order and then cooling in a tiny magnetic field as small as 0.1~mT to set the final octupole orientation. Our experiments on bulk single crystals give a clear baseline for this mechanism, free from the complexities of thin-film spin-torque effects. We show that the threshold field vanishes at $T_N$, and that crossing $T_N$ is essential for the weak field to control the final state. Based on these results, we argue that thermal softening (the temporary removal of magnetic anisotropy by Joule heating) is not a secondary effect but a key partner to spin–orbit torque in switching Mn$_3$Sn. In thin-film devices, the SOT effective field acts like the tiny cooling field, but switching cannot happen without enough heating to remove the anisotropy barrier. This unified picture helps explain earlier results and provides a rational basis for device design. We further provide a simple thermal model to estimate the temperature rise in nanoscale device cores, helping researchers predict whether a given current pulse will cause thermal softening. Finally, we suggest that ignoring thermal effects—or putting them aside—risks misunderstanding the switching mechanism and may lead to efforts that focus too much on spin-torque efficiency while overlooking thermal management. We hope this work encourages the community to adopt a more balanced view: in chiral antiferromagnets, heat and spin work together, not against each other.

\section*{Methods}
\subsection*{Sample preparation}
High-quality Mn$_3$Sn single crystals were grown by the Bridgman–Stockbarger method. The crystals were oriented by Laue diffraction and cut into bars of dimensions approximately $3 \times 1 \times 0.5$ mm$^3$ for Hall measurements. For the pulsed thermal annealing experiments, the sample was mounted on a resistive heater chip using insulating glue to avoid heating current passing through the sample. An E-type thermocouple was attached directly to the sample to monitor temperature, and a copper cylinder served as a heat sink to facilitate rapid cooling.

\subsection*{Measurements}
Hall resistivity $\rho_{zy}$ was measured using a standard four-probe method with a $\pm 10$~mA square-wave current to eliminate thermoelectric offsets. The heating current (22.2–24 mA, duration 10~s) was supplied by a Keithley 6221, and Hall voltages and thermocouple readings were recorded with a Keithley 2182A nanovoltmeter. All measurements were performed in a Physical Property Measurement System (PPMS, Quantum Design) with a base temperature of 250~K, where the AHE signal is large and stable. The sample temperature was controlled by the resistive heater, and the external magnetic field was applied in-plane along the $x$ direction.

\section*{Funding Declaration}
This work was supported by The National Key Research and Development Program of China (Grant No. 2023YFA1609600, 2024YFA1611200 and 2022YFA1403500), the National Science Foundation of China (Grant No. 12304065, 51821005, 12004123, 51861135104 and 11574097), the Fundamental Research Funds for the Central Universities (Grant No. 2019kfyXMBZ071), the Hubei Provincial Natural Science Foundation (2025AFA072).

\section*{Author contributions statement}
X.L. conceived and designed the study. X.L. grew the single crystal samples. J.Z. performed the measurements under the guidance of X.G. X.L. and Z.Z. X.L., J.Z. and Z.Z. analyzed the data. X.L. wrote the manuscript with input from all authors.

\section*{Competing interests}
The authors declare no competing financial or non-financial interests.

\section*{Data availability}
The main data supporting the findings of this study are available in the Article and its Supplementary Materials. Additional data are available from the corresponding authors on request.

\bibliography{main}

\end{document}